\documentclass{ifacconf}

\usepackage{graphicx}      % include this line if your document contains figures
\usepackage{graphics} % for pdf, bitmapped graphics files
\usepackage{epsfig} % for postscript graphics files
\usepackage{natbib}        % required for bibliography
\usepackage{amsmath}
\usepackage{amsfonts}
\usepackage{amssymb}  % assumes amsmath packet installed
\usepackage{bm}
\usepackage{enumitem} %to add letters for enumerations (eg A1, A2)
\usepackage{mathtools} %for \coloneq
\newcommand{\norm}[1]{\left\lVert#1\right\rVert}
\newtheorem{assumption}{{Assumption}}[section]

\graphicspath{ {./Pics/} }

\usepackage{eso-pic}

\newcommand{\publishernotetext}{%
	\textcopyright{} 2026 the authors. This work has been accepted to IFAC
	for publication under a Creative Commons Licence CC-BY-NC-ND.%
}

\newcommand{\copyrightnotice}{%
	\AddToShipoutPictureFG*{%
		\AtPageLowerLeft{%
			\raisebox{7mm}{%
				\makebox[\paperwidth][c]{%
					\fcolorbox{black}{white}{%
						\parbox{0.842\paperwidth}{%
							\centering\footnotesize
							\publishernotetext
						}%
					}%
				}%
			}%
		}%
	}%
}
    
\makeatletter
\let\old@ssect\@ssect % Store how ifacconf defines \@ssect
\makeatother

\usepackage[hidelinks,bookmarks=false]{hyperref}

\makeatletter
\def\@ssect#1#2#3#4#5#6{%
  \NR@gettitle{#6}% Insert key \nameref title grab
  \old@ssect{#1}{#2}{#3}{#4}{#5}{#6}% Restore ifacconf's \@ssect
}
\makeatother

\begin{document}
	\copyrightnotice
\begin{frontmatter}

\title{Optimal Delay Compensation in Networked Predictive Control} 

\thanks[footnoteinfo]{The authors would like to thank the Federal Ministry of Research, Technology, and Space (BMFTR) for its support as part of the research program Communication Systems “Souverän. Digital. Vernetzt.”. Joint project 6G-life, project identification number: 16KIS2414}
%\thanks[copyrightinfo]{\textcopyright{} 2026 the authors. This work has been accepted to IFAC for publication under a Creative Commons Licence CC-BY-NC-ND.}

\author[First]{Severin Beger} 
\author[First]{Yihui Lin} 
\author[Second]{Katarina Stanojevic}
\author[First]{Sandra Hirche}

\address[First]{Technical University of Munich (TUM), Munich, 80333 Germany (e-mail: \{severin.beger\}, \{yihui.lin\}, \{hirche\}@tum.de).}
\address[Second]{Technical University of Graz, Graz, 8010 Österreich (e-mail: katarina.stanojevic@tugraz.at)}

\begin{abstract}
Networked Predictive Control is widely used to mitigate the effect of delays and dropouts in Networked Control Systems, particularly when these exceed the sampling time. A key design choice of these methods is the delay bound, which determines the prediction horizon and the robustness to information loss. This work develops a systematic method to approximate the optimal delay bound with respect to a known delay distribution by quantifying the trade-off between prediction errors and open-loop operation caused by communication losses. Simulation studies demonstrate the performance gains achieved with the optimal bound.
\end{abstract}


\end{frontmatter}
%===============================================================================

\section{Introduction}
Modern control architectures rely increasingly on remote controllers and computations performed away from the plant. Examples include learning-based controllers that require substantial computational resources, as well as mobile and aerial robotic systems that offload control to cloud and edge infrastructures. While such networked control systems (NCS) offer many advantages, e.g., additional computational power, reduced energy requirements, and flexible reconfiguration, they are subject to communication-induced constraints (cf. \cite{Zhang.2020}). In particular, control over wireless all-purpose networks, such as WiFi (cf. \cite{Pezzutto.2024}) or 5G, exhibits round-trip times (RTT) larger than the sampling period and non-negligible packet losses, which fundamentally challenge the control design. \\
Strategies using predictive control methods for delay and dropout compensation, commonly referred to as networked predictive control (NPC), have received considerable attention over the last two decades. In an early work  \mbox{\cite{Bemporad.1998}} established the idea of forwarding the input sequences of a remote model predictive controller (MPC) to the actuator in order to counteract large delays. Several  works focus on pure delay compensation through prediction, such as \cite{PoiLoonTang.2006} as well as \cite{G.P.Liu.2007}, while others explicitly address dropouts, as seen in \cite{Quevedo.2007} and \cite{Pezzutto.2022}. Furthermore, \cite{L.Grune.2009} pointed out the importance of \emph{prediction consistency}, i.e., ensuring that the controller’s state predictions are based on the same inputs that are ultimately applied at the plant, when considering both delays and dropouts. Violations of prediction consistency may lead to infeasibility of the underlying optimal control problem and, in the worst case, to instability. Approaches guaranteeing prediction consistency include event-triggered or structure-preserving formulations, as in \cite{P.Varutti.2011} and \cite{G.Pin.2011}. Additionally, the inherent 
robustness of NPC schemes has been investigated by \cite{Findeisen.2011b}.\\
Recently, the NPC paradigm has been extended to scenarios with completely unacknowledged (User Datagram Protocol (UDP)-like) communication in systems without time synchronization, considering time-varying delays and dropouts in both the uplink (plant-to-controller) and downlink (controller-to-plant). \cite{G.Pin.2021} ensure consistency through a move-blocking strategy, whereas \cite{Beger.2024} introduce a two-mode architecture that distinguishes nominal operation from recovery after detected inconsistencies. Although these methods differ in several aspects, they share a crucial mechanism: 
a fixed delay bound $\bar{\tau}$ is introduced as a design parameter for delay compensation. The controller predicts the plant state $\bar{\tau}$ steps ahead and computes control inputs for the corresponding future application time. Packets arriving later than $\bar{\tau}$ steps are treated as dropouts and discarded.\\
The choice of the delay bound $\bar{\tau}$ is typically made conservatively, based on known bounds on delay and successive dropouts, or on engineering heuristics; yet, it has a direct and significant impact on closed-loop performance. A large bound increases prediction errors due to long forward roll-outs, while a small bound induces frequent dropouts and extended recovery phases from prediction inconsistencies. To the best of our knowledge, existing NPC schemes do not provide a systematic method for selecting $\bar{\tau}$, despite its central role in the compensation mechanism.\\
The contribution of this paper is two-fold. 
First, we develop a performance metric that quantitatively captures the trade-off between prediction errors during nominal operation and recovery-related errors due to late or lost packets, explicitly as a function of the chosen delay bound and RTT distribution. Second, we show that this metric admits a minimizer, yielding an optimal delay bound $\bar{\tau}^*$ for a given discrete delay distribution. The result provides a principled method for selecting the delay bound used in prediction-consistent NPC. We illustrate the effectiveness of the proposed approach in simulations.\\
The remainder of this work is structured as follows. Section~\ref{sec:ProbStatement} introduces the setup and network assumptions. Section~\ref{sec:NPC} introduces the NPC strategy, while Sections~\ref{sec:ErrorTradeOff} and~\ref{sec:OptimalDelayBound} derive the error analysis and the resulting optimal delay-bound selection method. Simulation results are provided in Section~\ref{sec:Sim}, followed by concluding remarks in Section~\ref{sec:Conclusion}.

\section{Setup and Problem Statement}
\label{sec:ProbStatement}
\begin{figure}[t]
\centering
\includegraphics[width=.9\linewidth]{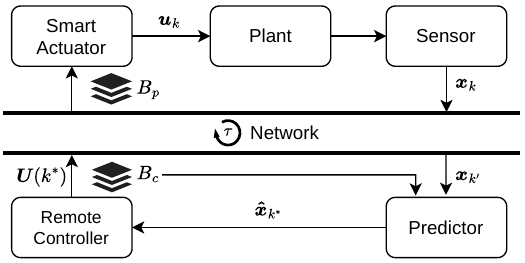}
\caption{Considered NPC setup over a lossy network.}
\label{fig:SystemOverview}
\end{figure}

Consider a plant with discrete-time, linear time-invariant (LTI) dynamics subject to additive noise 
\begin{align}
    \bm{x}_{k+1} = \bm{A}\bm{x}_k+\bm{B}\bm{u}_k + \bm{w}_k
    \label{eq:dynamics}
\end{align}
where $\bm{x}_k \in \mathbb{R}^n$ is the system state, $\bm{u}_k \in\mathbb{R}^m$ is the control input, and $\bm{w}_k \in \mathbb{W} \subseteq \mathbb{R}^n$ is an unknown, yet bounded additive disturbance with scalar bound $\bar{w}$. Furthermore, the plant is subject to state and input constraints, i.e., $\bm{x}_k \in \mathbb{X}, \bm{u}_k \in \mathbb{U}, \forall k$, with $\mathbb{U}, \mathbb{X}$ representing convex admissible sets. All admissible inputs in $\mathbb{U}$ are bounded by $\bar{u}$. Additionally, we assume full state feedback and the uniform sampling period $T_d$. 
\rem[Continuous Time] In the following, we only discuss time in a discrete sense. Continuous time can always be recovered by multiplying the discrete time steps with the sampling period, i.e., $t = k \cdot T_d$.\\
The system is connected to a remote controller over a non-acknowledged, UDP-like packet-based communication network, cf. Fig. \ref{fig:SystemOverview}.
All communication is subject to time-varying delays longer than a sampling step and packet losses. To characterize RTT delays $\tau \in \mathbb{N}$, all measurement packets from the sensor to the controller are time-stamped. The RTT  represents the time from issuing a measurement packet at the sensor up until a controller message based on said measurement arrives back at the plant. Thus, it aggregates delays from the sensor-to-controller, from the computation of new control values, and from controller-to-actuator. Furthermore, the following assumptions on the network hold:
\begin{assumption}	\label{as:delayIID}
	All RTT delays are independent and identically distributed (i.i.d.) according to the known discrete delay distribution $\mathcal{P}_\tau$, i.e. $p^\tau \sim \mathcal{P}_\tau$ with $p_k^\tau = \mathrm{Pr}(p^\tau= k), k \in \mathbb{N}$, $F(i)=\sum_{k=0}^i p^\tau_k$. 
\end{assumption}
\rem[On the i.i.d. Delay Assumption]{
While real\\communication networks can exhibit temporal correlation, e.g., congestion bursts or link-quality variations, this assumption is common in the NCS design and provides a tractable baseline for analysis. For the proposed strategy, the key requirement is  to compute the per-step dropout probability and the distribution of repeated failures. 
}
\begin{assumption}
	\label{as:synchronization}
The actuator and sensor are either collocated or synchronized and thus operate on a common clock. In contrast, the controller is not required to be synchronized with the plant side, preserving the flexibility of the networked architecture.
\end{assumption}  
\vspace{-1.5ex}
From Assumption~\ref{as:synchronization} it follows that the RTT can be determined using the time stamp information, whereas the individual delays within the loop remain unknown.\\
Within NPC, the delay bound $\bar{\tau}$ is a crucial design variable, as it determines how far the controller predicts ahead and, thereby, the maximum delay that can be compensated. The aim of this paper is to develop a systematic method for selecting $\bar{\tau}$ based on the RTT delay distribution. In the following, the considered NPC strategy is introduced, providing the structural and algorithmic framework required for the error analysis in Section~\ref{sec:ErrorTradeOff}.

\section{Networked Predictive Control Strategy}
\label{sec:NPC}
The core idea of NPC methods is to predict future states from past measurements and previously computed inputs with a constant prediction window of $\bar{\tau}$ steps, compute control sequences for these predicted states, and buffer them at the plant until their activation time arrives.
Thus, the delays, which are time-varying in reality, are rendered constant with chosen delay $\bar{\tau}$ and are compensated, at the cost of occasionally introducing artificial delay due to buffering. If values have RTT delays longer than $\bar{\tau}$, they are considered dropouts. Thus, with given delay probability distribution $\mathcal{P}_\tau$, the dropout probability for a given choice of $\bar{\tau}$ is defined as
\begin{align}
\label{eq:dropoutProbability}
    p_{d}(\bar{\tau}) = 1-F(\bar{\tau}).
\end{align}
Dropouts can be compensated for by sending input sequences rather than single values, and then buffering these sequences and iterating them at the plant, should the next sequence not arrive in time. This is known as sequence forwarding.
In this paper, we adopt the NPC method of \cite{Beger.2024} for delay and dropout compensation. Nevertheless, the reasoning of this work can be adapted to other known NPC methods, e.g. from \cite{L.Grune.2009} or \cite{G.Pin.2021}. In the following, we recall the elements required for the delay-bound analysis.\\
When a measurement $\bm{x}_{k'}$ taken at plant time~$k'$ arrives at the remote side, the controller predicts the state $\bm{\hat{x}}_{k^*}$ at  $k^*=k'+\bar{\tau}$. Using the nominal model and the control inputs stored in the buffer $B_c$ (cf. Fig. \ref{fig:SystemOverview}), the predicted state is
\begin{align}
\bm{\hat{x}_{k^*}} = \bm{A}^{\bar{\tau}}\bm{x}_{k'}+\sum_{i = 0}^{\bar{\tau}-1}\bm{A}^{\bar{\tau}-i-1}\bm{B}\hat{\bm{u}}_{k'+i}
\label{eq:predictedstate}
\end{align}
where $\hat{\bm{u}}_{k'+i} \in B_c$ denotes the  buffered input assumed to be applied at the plant at time $k'+i$.
Based on $\bm{\hat{x}}_{k^*}$, the controller computes an input sequence
 $\bm{U}(k^*) = [\bm{u}(0), \bm{u}(1), \dots, \bm{u}(N-1)]$
by solving the optimal control problem (OCP)
\begin{align}
\label{eq:OCP}
    \mathbb{P}(\bm{\hat{x}}_{k^*}) = \begin{cases}
        \min_{\bm{U}} & \hspace{-2mm}\sum_{i=0}^{N-1}\left(\bm{x}_i^T\bm{Q}\bm{x}_i + \bm{u}_i^T\bm{R}\bm{u}_i\right) + \bm{x}_N^T\bm{P}\bm{x}_N\\
      \textrm{s.t. } &\hspace{-2mm}\bm{x}_i \in \mathbb{X}, \bm{u}_i \in \mathbb{U}, \bm{x}_N \in \mathbb{X}_N,\\
         &\hspace{-2mm}\bm{x}_{i+1} = \bm{A}\bm{x}_i+\bm{B}\bm{u}_i,\\
        &\hspace{-2mm}\bm{x}_0 = \bm {\hat{x}}_{k^*}
    \end{cases}
\end{align}
where $N$ is the prediction horizon, $\bm{Q},\bm{P}\succeq 0$, $\bm{R}\succ 0$, and $\mathbb{X}_N$ is a terminal set associated with a stabilizing fallback control law $\bm{u}_k = \bm{\kappa}(\bm{x}_k)$. 
Finally, the resulting input sequence $ \bm{U}(k^*)$ is time-stamped for application from $k^*$ onward and transmitted to the plant, where it is stored in a plant buffer $B_p$ (cf. Fig. \ref{fig:SystemOverview}). 
At each plant time $k$, the actuator applies the first element of a valid, prediction-consistent sequence scheduled for $k$. If no such sequence is available, it advances the previously accepted sequence and applies $\bm{\kappa}(\bm{x}_k)$ once this sequence is exhausted.\\
 The reuse of buffered sequences creates the issue of prediction consistency, as introduced in \cite{L.Grune.2009}. If the controller predicts future states \eqref{eq:predictedstate} based on an input sequence the plant did not actually apply, the initial condition of the next OCP no longer represents the plant's evolution. 
 Following \cite{Beger.2024}, each input sequence therefore carries an identifier. A mismatch of identifiers indicates prediction inconsistency and triggers recovery.\\
Recovery consists of two plant-side modes: i) in the \emph{correction mode}, after discarding an inconsistent packet, the plant reuses the last consistent input sequence in open-loop until a correcting input sequence from the controller arrives; ii) in the \emph{acknowledgment mode}, the plant informs the controller of a successful error correction. Until this information reaches the controller, correction sequences may still be sent. Hence, the plant operates in three relevant modes - nominal operation, correction, and acknowledgment -, while the controller only knows a nominal and a correction mode.\\
For the mode analysis, we distinguish five combined plant/controller states: $(S_1)$:~Nominal operation; $(S_2)$:~Plant-side correction with nominal controller; $(S_3)$:~Correction at both sides; $(S_4)$:~Plant-side acknowledgment with correcting controller; $(S_5)$:~Plant-side acknowledgment with nominal controller.
\rem[Up-/Downlink Delay Distribution] {Since only the RTT distribution $\mathcal{P}_\tau$ is assumed to be known, we approximate uplink and downlink transmission success using~$\mathcal{P}_\tau$, inducing a geometric model for repeated transmission attempts. This neglects conditional dependencies between uplink and downlink delays but yields a tractable design rule. Distinct channel probabilities can be incorporated analogously. \label{rem:UpDownDelayDist}}
\vspace{1ex}\\
With dropout probability $p_d(\bar{\tau})$ as defined in \eqref{eq:dropoutProbability}, the transition probabilities between states can be described with a Markov chain as $\bm{\pi}_{k+1}=\bm{\pi}_k \bm{P}(p_{d})$
and its transition matrix
\begin{align}
    \label{eq:TransitionMatrix}
    \bm{P}(p_{d}) = \begin{bmatrix}
1-p_{d} & p_{d} &0 &0 &0\\
0 & p_{d} &1-p_{d} &0 &0\\ 
0 & 0 &p_{d} &1-p_{d} &0 \\
0 & 0 &0 &p_{d} &1-p_{d} \\
1-p_{d} & p_{d} &0 &0 &0
\end{bmatrix}.
\end{align}
The last row indicates that when the controller switches back from correction to nominal mode, the next message from the controller to the plant either causes the plant to return to nominal and thus overall to $(S_1)$, if transmission is successful, or it directly enters correction and thus $(S_2)$, if a dropout occurs. 
Solving for the stationary distribution \mbox{$\underset{k \to \infty}{\lim}\bm{\pi}_k=\bm{\pi}_k \bm{P}(p_{d})$} yields
\begin{equation}
\pi_{1} = \frac{(1-p_d)^2}{2p_d+1}, \,\, \pi_{2,3,4} = \frac{p_d}{2p_d+1}, \,\, \pi_{5} = \frac{p_d(1-p_d)}{2p_d+1}.
\label{eq:probratios}
\end{equation}
As the later error analysis is formulated in terms of plant-side modes, we define the plant-side mode weights as
\begin{align}
\label{eq:PhaseWeights}
\rho_{1}&=\pi_{1}, & 
\rho_{2}&=\pi_{2}+\pi_{3}, &
\rho_{3}&=\pi_{4}+\pi_{5},
\end{align}
representing the nominal, correction, and acknowledgment mode, respectively.\\
The choice of $\bar{\tau}$ directly influences how often the system operates in each of the modes described above. A large $\bar{\tau}$ reduces the likelihood of entering recovery but increases prediction errors due to longer forward roll-outs in \eqref{eq:predictedstate}. Conversely, a small $\bar{\tau}$ yields more frequent dropouts and thus causes transitions through the correction and acknowledgment modes, accumulating additional correction-related errors. Since the stationary mode ratios in \eqref{eq:probratios} determine how strongly each error type contributes to the long-term closed-loop behavior, identifying the optimal delay bound requires a quantitative characterization of the errors incurred in all modes, which we derive in the following section. 

\section{Error Trade{-}off in NPC}
\label{sec:ErrorTradeOff}

In this section, we derive scalar worst-case error bounds for the three plant-side modes introduced in Section \ref{sec:NPC}: nominal operation, correction, and acknowledgment. The bounds quantify deviations caused by prediction rollouts and by open-loop operation during recovery.\\
Let $\bm{e}_k = \bm{x}_k - \bm{\hat{x}}_k$ denote the mismatch between the actual and predicted state at the time of input application.
From~\eqref{eq:dynamics}, the error dynamics satisfy \begin{align}
\label{eq:errorDynamics}
    \bm{e}_{k+1} = \bm{A}\bm{e}_k + \bm{B}\Delta\bm{u}_k+\bm{w}_k
\end{align}
where 
    $\Delta\bm{u}_k = \bm{u}_k-\hat{\bm{u}}_k$
represents the mismatch between the applied input $\bm{u}_k$ and the input $\hat{\bm{u}}_k$ assumed in prediction \eqref{eq:predictedstate}. 
In the following, $\epsilon$ denotes scalar error values, and all norms are induced 2-norms.

\subsection{Fundamental Errors: Rollouts and Open-Loop-Behavior}
\label{sec:fundamentalErrors}

All three mode error bounds, $\epsilon_n(\bar{\tau})$ for the nominal mode, and $\epsilon_c(\bar{\tau})$ and $\epsilon_a(\bar{\tau})$ for correction and acknowledgment modes, respectively, originate from two fundamental sources:
(i) roll-out errors, caused by predicting future states from delayed measurements, and
(ii) open-loop errors, arising when the plant must temporarily reuse previously computed input sequences. Firstly, under prediction consistency, the inputs used in the rollout coincide with those applied at the plant, i.e. $\Delta\bm{u}_k=\bm{0}$. Hence, $\norm{\bm{e}_{k+1}}\leq \norm{\bm{A}}\norm{\bm{e}_{k}} + \bar{w} $, which yields the $l$-step rollout error bound
\begin{align}
\label{eq:rollout}
    E_r(l, \epsilon_0) = \norm{\bm{A}}^l \epsilon_0 + \bar{w}\sum_{i=0}^{l-1} \norm{\bm{A}^i},
\end{align}
where $\epsilon_0 = \norm{\bm{e}_0}$. Secondly, open-loop errors occur when the plant must reuse an input sequence multiple times. This occurs when either no new input sequence is sent for a specific actuation time or when a dropout and subsequent prediction inconsistency result in an error recovery scenario. During sequence reuse, the input mismatch can be bounded for linear MPC by a Lipschitz constant $L$ (cf. \cite{Teichrib.2023}) as $\norm{\Delta\bm{u}_k} \leq L \norm{\bm{e}_k} $. Thus,  $\norm{\bm{e}_{k+1}}\leq {\Lambda}\norm{\bm{e}_k} + \bar{w}$ with $\Lambda = {\norm{\bm{A}}+\norm{\bm{B}}L}$, and, as long as the buffered sequence contains inputs, i.e. $l < N$, we have the $l$-step open-loop error
\begin{align}
\nonumber
E_o^{(1)}(l,\epsilon_0) = \Lambda^l \epsilon_0 + \bar{w}\sum_{i=0}^{l-1}\Lambda^i,\quad \forall l< N.
\end{align}
If the input sequence is exhausted, the fallback input satisfies $\norm{\bm{B}(\bm{\kappa}(\bm{x}_k)-\hat{\bm{u}}_k)}\leq 2 \norm{\bm{B}}\bar{u}$, which gives
\begin{align}
\nonumber
E_o^{(2)}(l,e_0) = &\norm{\bm{A}^{l-N}} E_o^{(1)}(N-1,e_0)\\
&\nonumber+ (\bar{w}+2\norm{\bm{B}}\bar{u})\sum_{i=0}^{l-N}\norm{\bm{A}^i}, \quad\forall l\geq N.
\end{align}
Combined, these two error components give the piecewise open-loop error 
\begin{align}
\label{eq:OpenLoopComplete}
    E_o(l,e_0) = \begin{cases}
        E_o^{(1)}(l,e_0) &\mathrm{if }\,l <N\\
        E_o^{(2)}(l,e_0) &\mathrm{if }\, l\geq N.
    \end{cases} 
\end{align}
With \eqref{eq:rollout} and \eqref{eq:OpenLoopComplete}, we may now define the  resulting errors of each plant side mode. We will drop the initial error bound~$\epsilon_0$ in the following for brevity whenever it is zero.
\subsection{Expected Error in Nominal Mode}
\label{sec:NominalError}

In nominal operation, the plant always exhibits the rollout error $E_r(\bar{\tau})$ due to the prediction of the future state in NPC. Additional open-loop operation may occur due to packet disorder. Let $T \in \mathbb{N}$ be a random variable denoting the number of additional steps for which the plant must reuse an old input sequence due to packet disorder. Then, 
$\mathbb{E}(E_o(T,E_r(\bar{\tau})) = \sum_{i=0}^\infty \mathrm{Pr}(T=i)E_o(i,E_r(\bar{\tau}))$
holds by the law of total expectation. We compute $\mathrm{Pr}(T=i)$ from two disorder events. 
Firstly, an extension of $i$ steps occurs if the next packet arrives $i$ steps \emph{earlier} than the previous one. This can only occur for $i \leq\bar{\tau}-1$ steps. Let $X_1, X_2 \sim \mathcal{P}_\tau$ denote independent packet delays. Then $\mathrm{Pr}(T=i) = \mathrm{Pr}(X_2-~X_1 = -i) = p_{\Delta-}(i,\bar{\tau})$ holds. This can be explicitly computed as
\begin{align}
\label{eq:pEarlyArrival}
    p_{\Delta-}(i,\bar{\tau})&= \begin{cases}
        \sum_{k=0}^{\bar{\tau}-i} p^\tau_{k+i}\,p^\tau_k , &\text{if }\, i \in \{1,\dots,\bar{\tau}-1\}\\
        0 &\text{else}.
    \end{cases}
\end{align}
Secondly, an extension of length $i$ due to \emph{late} or \emph{missing} packets occurs when a measurement packet with delay $k$ is currently the most recent (probability $p^f_k$ as in \eqref{eq:pFreshest}) and no newer packet arrives in the next $i$ steps, yielding
    \begin{align}
\label{eq:pLateArrival}
    p_{\Delta+}(i,\bar{\tau})&= \sum_{k=0}^{\bar{\tau}}\bigg(p^f_k\prod_{j=k}^{k+i-1}(1-F(j))\bigg),\, i \in \mathbb{N}_{>0} \\
\label{eq:pFreshest}
    \text{and}\qquad p^f_k &=\frac{p^\tau_k\prod_{r=0}^{k-1}(1-p^\tau_r)}{1-\prod_{r=0}^{\bar{\tau}}(1-p^\tau_r)} .
\end{align}
Finally, the expected nominal error is therefore
\begin{align}
    \label{eq:PredictionError}
    \epsilon_n(\bar{\tau}) = &E_r(\bar{\tau}) +\nonumber\\ + &\sum_{i=1}^{\infty} \bigg( p_{\Delta-}(i,\bar{\tau})+p_{\Delta+}(i,\bar{\tau})\bigg) E_o(i,E_r(\bar{\tau})).
\end{align}
As the sum of the probabilities of both disorder events individually is $\leq 1$, \eqref{eq:PredictionError} is absolutely convergent. 

\subsection{Expected Errors in Recovery Mode}
When an inconsistency is detected at the plant, it reuses the last consistent input sequence until a correcting one arrives. During this time, the plant is in open-loop, and respective errors accumulate.  Let $T_c \in \mathbb{N}_0$ be the number of additional steps required to successfully correct a prediction inconsistency beyond the nominal RTT bound $\bar{\tau}$. Define  $\epsilon_o(\bar{\tau}) = E_o(\bar{\tau},\epsilon_n(\bar{\tau}))$ as the guaranteed error for one correction cycle, and $\Delta E_o(i,\bar{\tau})=E_o (\bar{\tau}+i,\epsilon_n(\bar{\tau}))-\epsilon_o(\bar{\tau})$ as the error for every additional step in the correction phase. Assuming RTT-based success probabilities as in Remark \ref{rem:UpDownDelayDist}, the expected per-step error for correcting a prediction inconsistency satisfies
\begin{align}
    \label{eq:CorrectionError}
    \epsilon_{c}(\bar{\tau}) = \frac{\epsilon_o(\bar{\tau})}{\bar{\tau}}+\sum_{i=1}^{\infty}p_c(i)\Delta E_o(i,\bar{\tau}) \qquad 
\end{align}
\begin{equation}
   \label{eq:pCorrection}
   \text{with} \quad p_c(i)=\mathrm{Pr}(T_c=i)= (1-p_{d}(\bar{\tau}))^2 (i+1)p_{d}(\bar{\tau})^i.
\end{equation}
Here, $p_c(i)$ follows from a negative-binomial distribution with two required successful transmissions, corresponding to one successful uplink and one successful downlink attempt.
Indeed, if $X_1,X_2 \sim\mathrm{Geom}(1-p_{d}(\bar{\tau}))-1$ denote the failed attempts before successful uplink and downlink transmission, then $T_c= X_1 + X_2$, and \eqref{eq:pCorrection} follows. 
The additional error from the pending correction for $i$ steps is given by $\Delta E_o(i,\bar{\tau})$. Applying the law of total expectation,
   $\epsilon_c(\bar{\tau})=\mathbb{E}[E_o(\bar{\tau}+T_c)]$
yields \eqref{eq:CorrectionError}.\\
During acknowledgment, only the successful uplink notification is required. Let $T_a \in \mathbb{N}_0$ be the number of additional steps required to successfully acknowledge the arrival of a correcting input sequence, such that the controller returns to nominal behavior. Assume that an acknowledgment requires at least one full RTT $\bar{\tau}$. Using the same $\epsilon_{o}(\bar{\tau})$ and $\Delta E_o(i,\bar{\tau})$, the expected per-step error for acknowledging the correction of a prediction error satisfies
\begin{align}
    \label{eq:AckError}
    \epsilon_{a}(\bar{\tau}) = \frac{\epsilon_o(\bar{\tau})}{\bar{\tau}}+\sum_{i=1}^{\infty}p_a(i)\Delta E_o(i,\bar{\tau})  
\end{align}
\begin{equation}
\label{eq:pAck}
\text{with} \quad p_a(i)=\mathrm{Pr}(T_a=i)= (1-p_{d}(\bar{\tau})) p_{d}(\bar{\tau})^i.
\end{equation}
The probability $p_a(i)$ is simply geometric, since the acknowledgment requires only one successful uplink transmission. Again, we may apply the law of total expectation, yielding \eqref{eq:AckError}.
\rem[Infinite Dropouts]
{\label{rem:InfiniteDropouts}Since the tail probabilities decay geometrically with $p_d(\bar{\tau})$, the infinite sums in \eqref{eq:PredictionError}, \eqref{eq:CorrectionError}, and \eqref{eq:AckError} can be truncated after $c_{\bar{\tau}}\in \mathbb{N}$ terms such that $(p_d(\bar{\tau}))^{\bar{\tau}+c_{\bar{\tau}}} < \phi$, where $0<\phi\ll1$. This ensures a bounded truncation error.} \\
\section{Optimal Delay Bounds in NPC}
\label{sec:OptimalDelayBound}
\begin{figure}[t]
\centering
\includegraphics{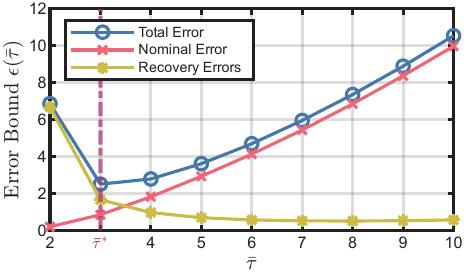}
\caption{The weighted errors in \eqref{eq:PerfIdx} illustrating  the trade-off between nominal and recovery errors. 
}
\label{fig:ErrorComparison}
\end{figure}

Combining these errors weighted by their expected fraction of time spent in each mode through the weights \eqref{eq:PhaseWeights} gives the performance index 
\begin{align}
\label{eq:PerfIdx}
\epsilon(\bar{\tau}) & =  \rho_1(\bar{\tau})\epsilon_n(\bar{\tau}) + \rho_2(\bar{\tau})\epsilon_{c}(\bar{\tau}) +\rho_3(\bar{\tau})\epsilon_{a}(\bar{\tau}),
\end{align}

where $\epsilon_n(\bar{\tau})$ represents the error in the nominal mode weighted with its total runtime ratio $\rho_1(\bar{\tau})$, while the error caused by the recovery mode is split into the correction phase $\epsilon_c(\bar{\tau})$ and the acknowledgment phase $\epsilon_a(\bar{\tau})$ with their respective ratios of the total runtime $\rho_2(\bar{\tau})$ and $\rho_3(\bar{\tau})$.\\
This metric, in turn, can be minimized to find the optimal delay bound choice 
\begin{align}
\label{eq:optimalBound}
    \bar{\tau}^* = \underset{\bar{\tau}}{\mathrm{argmin}} \,\epsilon(\bar{\tau}).
\end{align}
For each $\bar{\tau} \in \mathbb{N}$, the performance index $\epsilon(\bar{\tau})$ is well-defined, non-negative, and finite. Since the optimization domain is discrete and bounded by construction,  $\bar{\tau}^*$ always exists. In practice, $\bar{\tau}^*$ is obtained by evaluating \eqref{eq:PerfIdx} over the finite set of admissible delay bounds and selecting the minimizer. The infinite sums are truncated according to Remark \ref{rem:InfiniteDropouts}. Hence, the method does not require modifying the underlying NPC controller. It only provides an offline, or repeatedly updated online, design rule for the delay bound used by the existing prediction and buffering mechanism.\\
The derived performance metric in \eqref{eq:PerfIdx} captures the trade-off between the expected errors occurring during nominal operation and those arising during recovery with respect to the chosen delay bound $\bar{\tau}$.
Because $\epsilon_n(\bar{\tau})$ grows with increasing $\bar{\tau}$, while $\epsilon_c(\bar{\tau})$ and $\epsilon_a(\bar{\tau})$ increase as $\bar{\tau}$ decreases, the resulting performance index exhibits a unique minimum in all evaluated scenarios. Fig. \ref{fig:ErrorComparison} shows an exemplary sweep of \eqref{eq:PerfIdx} over a range of RTT delay bounds and highlights the discussed error trade-off.\\
Furthermore, $\bar{w}$ amplifies both rollout and open-loop errors and thus, a larger disturbance bound increases the penalty on long prediction horizons. Therefore, larger disturbances typically shift the optimal delay bound toward smaller values.\\
In the next section, we demonstrate that considering the optimal delay bound~$\bar{\tau}^*$ yields higher performance than selecting an arbitrary constant value.

\section{Simulation Results}
\label{sec:Sim}
The performance of the proposed strategy is demonstrated in the simulation study, considering a mass-spring-damper system given by
\begin{align}
\label{eq:exampleSys}
    \dot{\bm{x}} = \begin{bmatrix}
        0 &1\\
        -k/m &-d/m
    \end{bmatrix} \bm{x} + \begin{bmatrix}
        0 \\1/m
    \end{bmatrix} \bm{u} + \bm{w}_k,
\end{align}

with $m=1\mathrm{ kg}$, $k=10\, {\mathrm{N}}/{\mathrm{m}}$ and $d=0.5\,{\mathrm{Ns}}/{\mathrm{m}}$. The states represent position $x_1$ and velocity $x_2$, respectively. We choose a sampling time $T_d = 50\mathrm{ms}$ for discretisation and add white noise to the acceleration bounded by $\bar{w} = 0.1 \frac{\mathrm{N}}{\mathrm{kg}}$. The chosen second-order system has a fairly low damping ratio of $\zeta ={d}/({2\sqrt{km}})\approx0.79$ and a natural frequency of $\omega_0 = \sqrt{{k}/{m}}\approx 3.16 \mathrm{\frac{1}{s}}$. In conjunction with the rather long sampling time, this makes the system sensitive to noise and thus to state errors due to open-loop behavior caused by communication imperfections. 
The inputs are constrained as $\lvert\bar{u}\rvert=25 \mathrm{N}$. With MPC weighting matrices $\bm{Q} = \mathrm{diag}(500,1)$ and $\bm{R}=0.1$, and a time horizon $N=10$ we can approximate the global Lipschitz bound as $L=87.3$ through an explicit MPC method. All simulations\footnote{The code is available at \href{https://github.com/TUM-ITR/OptDelComp}%
{\texttt{github.com/TUM-ITR/OptDelComp}}.} are carried out in \emph{Matlab/Simulink 2025b}, using the toolbox \emph{NetFlex} from \cite{Stanojevic.2025}. \\
\begin{figure}[t]
\centering
\includegraphics{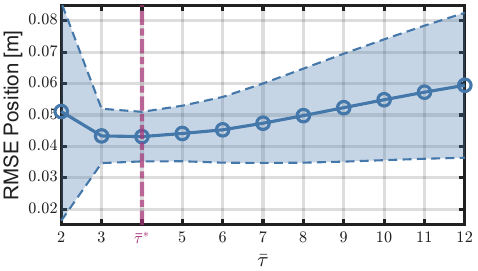}
\caption{A sweep of the position RSME over several delays for the presented algorithm (100 runs). The optimum aligns with the predicted value $\bar{\tau}^*=4$. }
\label{fig:DelaySweep}
\end{figure}
We sample all delays from log-normal distributions with parameters $\mu$ and $\sigma$, and subsequently discretize them. These distributions possess heavy tails and resemble typical network delay behavior. 
\begin{figure}[b]
\centering
\includegraphics[width=\linewidth]{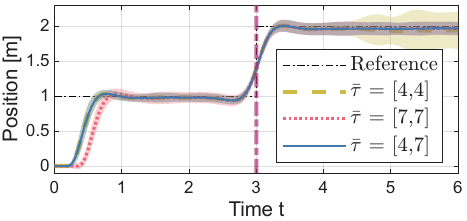}
\caption{Comparison of three cases with different prediction windows. The delay  distribution changes at $t=3s$ such that $\bar{\tau}^*_1=4$ and $\bar{\tau}^*_2=7$.  Using an adaptive delay bound (blue line) results in the highest performance.}
\label{fig:StateSweep}
\end{figure}
We then allocate the delay between the sensor–to-controller and controller–to-actuator channels by sampling a random split proportion $p_s$ drawn from a beta distribution with parameters $\alpha=\beta=2$ and setting $\tau_{sc,k}= \mathrm{round}(p_s\tau_k)$ and $\tau_{ca,k}=\tau_k-\tau_{sc,k}$. This ensures that the simulated RTTs match the theoretical distribution, while allowing for variability and asymmetry between uplink and downlink delays.\\
At first, we consider the delay distribution $\mathcal{P}^1_\tau$ as a log-normal distribution with $\mu_1=0.5,\sigma_1=0.5$, resulting in a mean delay of $3.1$. Evaluating \eqref{eq:optimalBound} leads to $\bar{\tau}^*=4$. To evaluate performance, we apply two unit steps to the position reference at $t=0$s and $t=3$s, respectively, and conduct 100 simulation runs. Each run sweeps over a set of potential delay bounds using the same delay distribution. The resulting RMSE position errors with their $95$\% confidence interval are depicted in Fig. \ref{fig:DelaySweep}. We observe that the computed optimal delay bound lines up with the lowest average error, although adjacent values are not far off. This is to be expected, as the optimization minimizes the expected worst-case error. Importantly, the proposed bound is optimal in expectation and demonstrates the highest consistency, as seen by the minimal variance around its optimum. The flatness of the RMSE curve around the optimum also indicates that the proposed metric identifies a robust operating region, rather than a fragile point optimum depending on a particular delay realization.\\
In a second experiment with the same reference, we alter the delay distribution at $t=3$s from $\mathcal{P}^1_\tau$ to $\mathcal{P}^2_\tau$ with $\mu_2=1.5$ and $\sigma_2=0.5$. For each distribution, we compute the optimal time delay bound as $\bar{\tau}_1^*= 4$ and $\bar{\tau}_2^*=7$, respectively. To assess performance, we conduct three different cases: the first two apply one of the optimal delay bounds for the entire duration, while the third run adapts the delay bound at the switching time, thereby utilizing the optimal delay bound at all times. We conducted this experiment 100 times, each with varying disturbances and resampled delays. Fig. \ref{fig:StateSweep} shows the resulting position over time, together with a $95$\% confidence interval depicted as shaded regions. The result indicates that the adaptive strategy outperforms the constant ones, suggesting that the optimal delay bound choice indeed leads to a performance increase and may be valuable as an adaptive strategy for online sampled delay distributions.

\section{Conclusion} 
\label{sec:Conclusion}
This work introduces a systematic method for approximating the optimal delay bound $\bar{\tau}^*$ in networked predictive control based on a known delay distribution $\mathcal{P}_\tau$. By decomposing the closed-loop behavior into nominal, correction, and acknowledgment modes, we derived a performance metric \eqref{eq:PerfIdx} that captures the fundamental trade-off between prediction-induced errors and open-loop behavior caused by packet dropouts. Minimizing  \eqref{eq:PerfIdx} yields the delay bound that optimizes expected closed-loop performance.
In simulations, we have shown the effectiveness of the method. The proposed framework provides a principled design tool for NPC schemes and establishes a foundation for future work on correlated or time-varying delay models and online adaptation of the delay bound.

\section*{DECLARATION OF GENERATIVE AI AND AI-ASSISTED TECHNOLOGIES IN THE WRITING PROCESS}
During the preparation of this work, the authors utilized ChatGPT and Grammarly to enhance the writing and technical content. After using these tools, the authors reviewed and edited the content as needed and take full responsibility for the content of the publication.

\bibliography{IFACWC2026_Beger}  

@inproceedings{Beger.2024,
 author = {Beger, S. and Hirche, S.},
 title = {{A Robust Model Predictive Control Method for Networked Control Systems}},
 url = {https://www.scopus.com/inward/record.uri?eid=2-s2.0-86000636622&doi=10.1109%2fCDC56724.2024.10886340&partnerID=40&md5=fb22846734cbdfebc51001da28510f24},
 pages = {6896--6903},
 isbn = {2576-2370},
 booktitle = {{2024 IEEE 63rd Conference on Decision and Control (CDC)}},
 year = {2024},
 doi = {10.1109/CDC56724.2024.10886340}
}

@inproceedings{Teichrib.2023,
 author = {Teichrib, D. and Darup, M. S.},
 title = {{Efficient Computation of Lipschitz Constants for MPC with Symmetries}},
 url = {https://www.scopus.com/inward/record.uri?eid=2-s2.0-85184796192&doi=10.1109%2fCDC49753.2023.10383472&partnerID=40&md5=085a024c8de0b3cdb64d8aaf2f871f46},
 pages = {6685--6691},
 booktitle = {{2023 IEEE Conference on Decision and Control (CDC)}},
 year = {2023},
 doi = {10.1109/CDC49753.2023.10383472}
}

@inproceedings{Stanojevic.2025,
 author = {Stanojevic, K. and Steinberger, M. and Siljak, M. and Ludwiger, J. and Horn, M.},
 title = {{NetFlex: A Simulation Framework for Networked Control Systems}},
 booktitle = {{IFAC 11th International Conference on Control, Decision and Information Technologies (CoDIT) 2025}},
 year = {2025}
}

@article{Pezzutto.2024,
 author = {Pezzutto, M. and Dey, S. and Garone, E. and Gatsis, K. and Johansson, K. H. and Schenato, L.},
 year = {2024},
 title = {{Wireless control: Retrospective and open vistas}},
 url = {https://www.sciencedirect.com/science/article/pii/S1367578824000403},
 pages = {100972},
 volume = {58},
 issn = {1367-5788},
 journal = {{Annual Reviews in Control}},
 doi = {10.1016/j.arcontrol.2024.100972}
}

@article{Zhang.2020,
 author = {Zhang, X.-M. and Han, Q.-L. and Ge, X. and Ding, D. and Ding, L. and {Yue. D.} and Peng, C.},
 year = {2020},
 title = {{Networked control systems: A survey of trends and techniques}},
 url = {https://www.scopus.com/inward/record.uri?eid=2-s2.0-85069913710&doi=10.1109%2fJAS.2019.1911651&partnerID=40&md5=5e731e1521aa2e3bca7dc8f416ae8829},
 pages = {1--17},
 volume = {7},
 number = {1},
 journal = {{IEEE/CAA Journal of Automatica Sinica}},
 doi = {10.1109/JAS.2019.1911651}
}

@inproceedings{Bemporad.1998,
 author = {Bemporad, A.},
 title = {{Predictive control of teleoperated constrained systems with unbounded communication delays}},
 pages = {2133-2138 vol.2},
 isbn = {0191-2216},
 booktitle = {{Proceedings of the 37th IEEE Conference on Decision and Control}},
 year = {1998},
 doi = {10.1109/CDC.1998.758651}
}

@article{Findeisen.2011b,
 author = {Findeisen, R. and Gr{\"u}ne, L. and Pannek, J. and Varutti, P.},
 year = {2011},
 title = {{Robustness of Prediction Based Delay Compensation for Nonlinear Systems}},
 url = {https://www.sciencedirect.com/science/article/pii/S1474667016436116},
 pages = {203--208},
 volume = {44},
 number = {1},
 issn = {1474-6670},
 journal = {{IFAC Proceedings Volumes}},
 doi = {10.3182/20110828-6-IT-1002.03090}
}

@article{G.P.Liu.2007,
 author = {Liu, G.-P. and Xia, Y. and Chen, J. and Rees, D. and Hu, W.},
 year = {2007},
 title = {{Networked Predictive Control of Systems With Random Network Delays in Both Forward and Feedback Channels}},
 pages = {1282--1297},
 volume = {54},
 number = {3},
 issn = {1557-9948},
 journal = {{IEEE Transactions on Industrial Electronics}},
 doi = {10.1109/TIE.2007.893073}
}

@article{G.Pin.2011,
 author = {Pin, G. and Parisini, T.},
 year = {2011},
 title = {{Networked Predictive Control of Uncertain Constrained Nonlinear Systems: Recursive Feasibility and Input-to-State Stability Analysis}},
 pages = {72--87},
 volume = {56},
 number = {1},
 issn = {1558-2523},
 journal = {{IEEE Transactions on Automatic Control}},
 doi = {10.1109/TAC.2010.2051091}
}

@article{G.Pin.2021,
 author = {Pin, G. and Fenu, G. and Casagrande, V. and Zorzenon, D. and Parisini, T.},
 year = {2021},
 title = {{Robust Stabilization of a Class of Nonlinear Systems Controlled Over Communication Networks}},
 pages = {3036--3051},
 volume = {66},
 number = {7},
 issn = {1558-2523},
 journal = {{IEEE Transactions on Automatic Control}},
 doi = {10.1109/TAC.2020.3021039}
}

@inproceedings{L.Grune.2009,
 author = {Gr{\"u}ne, L. and Pannek, J. and Worthmann, K.},
 title = {{A prediction based control scheme for networked systems with delays and packet dropouts}},
 pages = {537--542},
 isbn = {0191-2216},
 booktitle = {{Proceedings of the 48h IEEE Conference on Decision and Control (CDC) held jointly with 2009 28th Chinese Control Conference}},
 year = {2009},
 doi = {10.1109/CDC.2009.5399922}
}

@inproceedings{P.Varutti.2011,
 author = {Varutti, P. and Findeisen, R.},
 title = {{Event-based NMPC for networked control systems over UDP-like communication channels}},
 pages = {3166--3171},
 isbn = {2378-5861},
 booktitle = {{Proceedings of the 2011 American Control Conference}},
 year = {2011},
 doi = {10.1109/ACC.2011.5991331}
}

@article{Pezzutto.2022,
 author = {Pezzutto, M. and Farina, M. and Carli, R. and Schenato, L.},
 year = {2022},
 title = {{Remote MPC for Tracking Over Lossy Networks}},
 pages = {1040--1045},
 volume = {6},
 journal = {{IEEE Control Systems Letters}},
 doi = {10.1109/LCSYS.2021.3088749}
}

@article{PoiLoonTang.2006,
 author = {Tang, P. L. and de Silva, C. W.},
 year = {2006},
 title = {{Compensation for transmission delays in an ethernet-based control network using variable-horizon predictive control}},
 pages = {707--718},
 volume = {14},
 number = {4},
 issn = {1558-0865},
 journal = {{IEEE Transactions on Control Systems Technology}},
 doi = {10.1109/TCST.2006.876640}
}

@inproceedings{Quevedo.2007,
 author = {Quevedo, D. and {Silva E.} and {Goodwin G.}},
 title = {{Packetized Predictive Control over Erasure Channels}},
 pages = {1003--1008},
 booktitle = {{2007 American Control Conference}},
 year = {2007},
 doi = {10.1109/ACC.2007.4282630}
}

\end{document}